\documentclass[english,twocolumn,journal]{IEEEtran}
\usepackage[T1]{fontenc}
\usepackage{amsthm}
\usepackage{amssymb}
\usepackage{graphicx}

\makeatletter
\theoremstyle{plain}
\newtheorem{thm}{\protect\theoremname}
\theoremstyle{remark}
\newtheorem{rem}[thm]{\protect\remarkname}

\IEEEoverridecommandlockouts
\usepackage{ifpdf}
\usepackage{cite}
\usepackage{amsmath}
\usepackage{float}
\usepackage{array}
\usepackage{makecell}
\usepackage{url}
\usepackage{changebar}

\makeatother

\usepackage{babel}
\providecommand{\remarkname}{Remark}
\providecommand{\theoremname}{Theorem}

\begin{document}

\title{Low-Complexity Downlink User Selection for Massive MIMO Systems}

\author{Haijing Liu, Hui Gao, \emph{Member, IEEE}, Shaoshi Yang, \emph{Member, IEEE},
and Tiejun Lv, \emph{Senior Member, IEEE}%
\thanks{This work is financially supported by the National Natural Science
Foundation of China (NSFC) (Grant No. 61271188, 61401041), the
Fundamental Research Funds for the Central Universities (Grant No.
2014RC0106), and Beijing Municipal Science and Technology Commission research fund project No. D151100000115002.
 
Haijing Liu, Hui Gao, and Tiejun Lv are with the Beijing University
of Posts and Telecommunications, Beijing, China 100876 (e-mail: \{Haijing\_LIU,
huigao, lvtiejun\}@bupt.edu.cn).

Shaoshi Yang is with the School of Electronics and Computer Science,
University of Southampton, SO17 1BJ Southampton, U.K. (e-mail: sy7g09@ecs.soton.ac.uk).%
}}

\markboth{Accepted to publish on IEEE Systems Journal -- Special Issue on 5G Wireless Systems with Massive MIMO, April 2015.}%
{Shell \MakeLowercase{\textit{et al.}}: Bare Demo of IEEEtran.cls
for Journals}

\maketitle
\begin{abstract}
In this paper we propose a pair of low-complexity user selection schemes
with zero-forcing precoding for multiuser massive MIMO downlink systems,
in which the base station is equipped with a large-scale antenna array.
First, we derive approximations of the ergodic sum rates of the systems
invoking the conventional random user selection (RUS) and the location-dependant
user selection (LUS). Then, the optimal number of simultaneously served
user equipments (UEs), $K^*$, is investigated to maximize the sum
rate approximations. Upon exploiting $K^*$, we develop two user selection
schemes, namely $K^*$-RUS and $K^*$-LUS, where $K^*$ UEs are selected
either randomly or based on their locations. Both of the proposed
schemes are independent of the instantaneous channel state information
of small-scale fading, therefore enjoying the same extremely-low computational
complexity as that of the conventional RUS scheme. Moreover, both
of our proposed schemes achieve significant sum rate improvement over
the conventional RUS. In addition, it is worth noting that like the
conventional RUS, the $K^*$-RUS achieves good fairness among UEs. \end{abstract}


\section{Introduction}

\renewcommand{\figurename}{Fig.}

The multiuser MIMO (MU-MIMO) technology plays a key role in modern
wireless communications due to its substantial performance gains over
the conventional single-input single-output (SISO) techniques\cite{3GPP_4,3GPP_3}.
Relying on MU-MIMO, a multi-antenna base station (BS) can simultaneously
serve multiple user equipments (UEs) within a cell using the same
spectrum resource, and thus the spectral efficiency is improved. User
selection is critical for optimizing MIMO systems' overall performance
in a variety of scenarios and has been extensively studied, such
as in cellular networks (see for example \cite{yoo_optimality_2006}
and references therein) and in multi-hop networks\cite{gao_new_2013,gao_multiuser_2013,gao_distributed_2014,liu_novel_2014}.
The semi-orthogonal user selection (SUS) leveraging the degree of
channel orthogonality among UEs is probably one of the most popular
low-complexity user selection methods for improving system sum rates\cite{yoo_optimality_2006,chen_resource_2013,cheung_spectral_2014}.
Additionally, considering the fairness amongst UEs, round robin scheduling\cite{yoo_optimality_2006}
and random user selection (RUS) are regarded as the two simplest methods
offering equal opportunities to all the candidate-UEs. Both of them
have been widely employed in practical cellular systems as well\cite{3GPP_4}. 

In the realm of MU-MIMO, the recently proposed massive MIMO, where
the BS is equipped with a large-scale antenna array to serve multiple
UEs, has been widely envisaged as one of the major candidate technologies
for the fifth generation (5G) cellular networks owing to its favorable
features, such as huge spectral efficiency and energy efficiency gains\cite{yang_performance_2013,rusek_scaling_2013,zhang_capacity_2013,gaohui_2014,hu_esprit-based_2014}.
Like in the conventional small-scale MU-MIMO systems, user selection
is also important in massive MU-MIMO systems\cite{bjornson_optimal_2014,nam_joint_2014,lee_asympt_2014,xu_user_2014},
though it faces new challenges. More specifically, in the user selection
for conventional MU-MIMO systems, it is usually assumed that the number
of candidate-UEs, $N$, is much larger than that of the BS antennas,
$M$. Therefore, upon employing instantaneous channel state information
(CSI)-aided user selection methods (e.g. SUS), multiuser diversity
gains can be harvested to boost the overall system performance. By
contrast, in massive MIMO systems, it is impractical to have $N\gg M$,
since $M$ is already very large. Moreover, the computational complexity
of the conventional user selection methods might be too high for the
massive MIMO systems. For example, the computational complexity of
SUS is roughly $\mathcal{O}(M^3N)$\cite{yoo_optimality_2006}, which
will cause huge consumption of power and computational resources if
$M$ becomes large.

Recently, a range of user selection schemes have been proposed for
massive MIMO systems. The time-division duplex (TDD) and frequency-division
duplex (FDD) based massive MIMO systems impose different requirements
on user selection. By exploiting the instantaneous CSI of candidate-UEs,
Lee et al. proposed an SUS-like user selection method in \cite{lee_asympt_2014}
and Xu et al. developed a greedy user selection scheme in \cite{xu_user_2014}.
These selection methods mainly focus on FDD scenarios, in which the
amount of downlink transmission resources consumed by the downlink
channel estimation training for all the candidate-UEs does not increase
with the number of candidate-UEs $N$\cite{kobayashi_training_2011}.
By contrast, in TDD scenarios, the downlink channel is estimated through
uplink training relying on channel reciprocity, and the pilot/training
symbol overhead imposed by channel estimation increases with $N$\cite{marzetta_how_2006}.
In this scenario, if the number of candidate-UEs is large, most of
the channel coherence slot in time domain will be consumed by channel
estimation, leaving only a small fraction for downlink data transmission.
Hence, besides the high computational complexity, the pilot overhead
needed for channel estimation of candidate-UEs also limits the application
of instantaneous CSI-aided user selection methods in TDD scenarios.
In addition, for FDD based massive MIMO systems, Nam et al. introduced
user selection methods based on the candidate-UEs' feedback of instantaneous
signal-to-interference-plus-noise ratio (SINR) in\cite{nam_joint_2014}.
However, these methods are not applicable for TDD systems either.
This is because in TDD scenarios, the feedback signals from a large
number of candidate-UEs will increase the uplink proportion of the
uplink-downlink shared frequency band at each coherence slot, resulting
in reduced resources left for downlink data transmission. In summary,
neither the instantaneous-CSI estimation based nor the uplink-feedback
based user selection methods are suitable for the downlink of TDD-based
massive MIMO systems. At the time of writing, the design of user selection
for massive MIMO systems with the TDD mode remains a largely open
area. Hence, the novel user selection methods which cause no or just
little decrease of downlink transmission resources represent a new
promising research subject.

In this paper, we consider the downlink of a TDD based massive MIMO
system where pilot-based channel estimation and zero-forcing (ZF)
precoding are invoked for serving a number of UEs. First, with the
aid of the random matrix theory (RMT)-based large system analysis,
we derive approximations of the ergodic sum rates of the systems invoking
the conventional RUS and the location-dependent user selection (LUS).
The optimal number of simultaneously served UEs, denoted as $K^*$,
is solved offline for maximizing the sum rate approximations. Then,
aiming for improving the system sum rates, a pair of $K^*$-based
low-complexity user selection methods are proposed, namely the $K^*$-based
random user selection ($K^*$-RUS) and the $K^*$-based location-dependant
user selection ($K^*$-LUS). For $K^*$-RUS, $K^*$ UEs are randomly
selected for simultaneous data transmissions at each time slot. The
system sum rates are improved with an appropriate configuration of
$K^*$. Meanwhile, the fairness among UEs is guaranteed as a result
of the random selection. For the $K^*$-LUS scheme, $K^*$ UEs nearest
to the BS are selected for data transmission, which may achieve higher
sum rate performance than $K^*$-RUS.

Notably, our schemes exhibit two fundamental differences as compared
with the conventional user selection schemes. First, unlike the conventional
SUS that requires the instantaneous CSI of small-scale fading (SSF),
our schemes only need long-term CSI. Second, rather than emphasizing
which UEs should be selected for improving system performance, the
proposed user selection schemes mainly focus on how many UEs should
be selected for simultaneous transmissions. Thanks to these differences,
we bypass the complicated online computations regarding the sum rates
and the selection metric, which are often inevitable in the conventional
schemes. Therefore, the online computational complexity of the proposed
two schemes is on the same order as that of the conventional RUS scheme.
Furthermore, since our user selection schemes are independent of the
SSF CSI of candidate-UEs, we no longer have to carry out channel estimation
of all the candidate-UEs for user selection. Instead, only the active-UEs
need to send pilots at each coherence slot. As a beneficial result,
we are capable of saving the cost of channel training significantly
and attaining more resources for data transmission.

It is worth pointing out that a location-adaptive transmission strategy
was proposed for TDD based massive MIMO systems in \cite{huh_phd_2012}.
Our work differs from \cite{huh_phd_2012} in several respects. Random
locations of UEs are assumed in this paper, whereas in \cite{huh_phd_2012}
the UEs were assumed to be placed at fixed points. Therefore, the
optimal number of active-UEs in our paper is independent of specific
channel realizations, while in \cite{huh_phd_2012} this number has
to be re-calculated whenever any candidate-UE's large-scale-fading
(LSF) CSI changes. Moreover, we consider spatial correlation in the
channel model, which is ignored in \cite{huh_phd_2012}.

The remainder of this paper is organized as follows. The system model
is presented in Section II. In Section III, we analyze the asymptotic
sum rates of the system invoking the conventional RUS and LUS, and
then further develop two low-complexity user selection schemes. Our
numerical results are provided in Section IV. Finally, the conclusions
are drawn in Section V.

\emph{Notations:} We use uppercase and lowercase boldface letters
to denote matrices and vectors, respectively. $(\cdot)^H$, $(\cdot)^\dagger$
and $\textnormal{tr}(\cdot)$ denote the conjugate transpose, the
pseudo-inverse, and the trace operations, respectively. $\text{E}_x[\cdot]$
represents the expected value with respect to $x$. $\mathcal{CN}(\mathbf{m},\mathbf{\Theta})$
denotes the circularly symmetric complex Gaussian distribution with
mean vector $\mathbf{m}$ and covariance matrix $\mathbf{\Theta}$.
Finally, $\xrightarrow{a.s.}$ denotes the \emph{almost sure} convergence.

\section{System Model \label{sec:System-Model}}

\begin{figure}[!t]
\begin{centering}
\includegraphics[width=6cm]{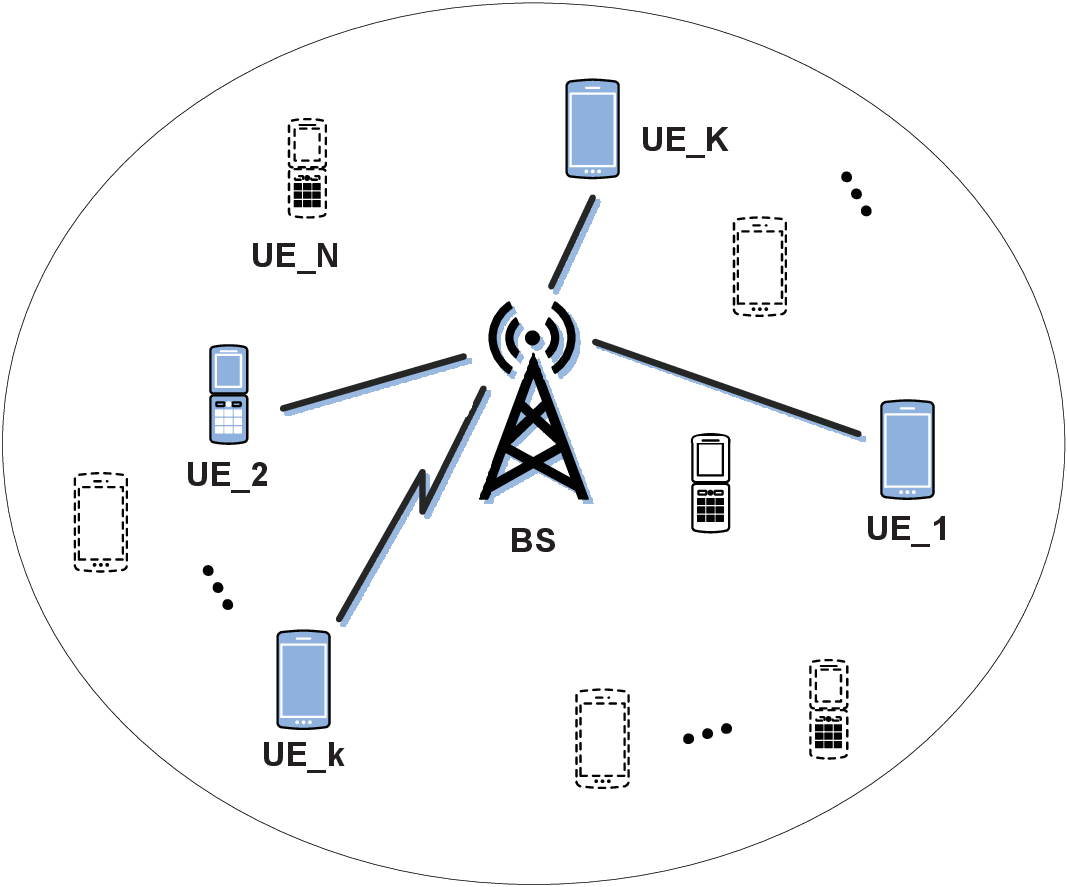}
\par\end{centering}

\protect\caption{The downlink of a TDD based massive MIMO system, which is composed
of an $M$-antenna BS and $N$ single-antenna candidate-UEs. Among
all the candidate-UEs, $K$ UEs are selected to be simultaneously
served, which are regarded as the active-UEs.}

\centering{}\label{FIG 1}
\end{figure}

We consider the downlink of a TDD based massive MIMO system consisting
of an $M$-antenna BS and $N$ single-antenna candidate-UEs ($N\geq M$).
We assume that $N$ and $M$ are of the same order. $K$ UEs $(K<M)$
are selected for simultaneous data transmissions at each coherence
slot. The composite channel matrix $\mathbf{G}\in\mathbb{C}^{K\times M}$
from the BS to the $K$ active-UEs characterizes LSF, SSF and transmit
correlation\footnote{The distance between UEs is supposed to be sufficiently large compared to the signal wavelength, so the receive correlation is not taken into account.},
and can be expressed as \begin{equation}\label{eq_g_0}\mathbf{G}=\mathbf{D}^{1/2}\mathbf{H}\mathbf{R}^{1/2},\end{equation}
where $\mathbf{H}\in\mathbb{C}^{K\times M}$ is the SSF matrix with
independent and identically distributed (i.i.d.) $\mathcal{CN}(0,1)$
entries, and the diagonal matrix $\mathbf{D}\in\mathbb{R}^{K\times K}$
contains the LSF coefficients $\beta_k$ along its main diagonal.
We model the LSF of the $k$-th active-UE as\footnote{In this paper, we use a simplified LSF model in which the shadow fading is excluded. Nevertheless, it should be noted that algorithms and schemes developed in this paper can be directly extended to the model including shadow fading.}
$\beta_k=cd_k^{-\alpha},k=1,\dots,K$, in which $d_k$ is the distance
from the $k$-th active-UE to the BS, $\alpha$ is the pathloss exponent
and $c$ is the pathloss at the reference distance. The transmit correlation
matrix at the BS is modeled by the widely used $\delta$-Kac-Murdock-Szeg?
matrix $\mathbf{R}$, in which $\delta$ is the antenna correlation
coefficient\cite{Zelst02asingle}.

The system operates in the TDD mode and the BS obtains the SSF CSI
relying on the training-based channel estimation. The estimation $\hat{\mathbf{H}}$
of the SSF CSI matrix $\mathbf{H}$ is modeled as \begin{equation}\label{eq_h_0}\mathbf{H}=\hat{\mathbf{H}}+\tilde{\mathbf{H}},\end{equation}where
the $K\times M$ dimensional estimated channel matrix $\hat{\mathbf{H}}=[\hat{\mathbf{h}}_1^T,\cdots,\hat{\mathbf{h}}_K^T]^T$
and the $K\times M$ dimensional error matrix $\tilde{\mathbf{H}}=[\tilde{\mathbf{h}}_1^T,\cdots,\tilde{\mathbf{h}}_K^T]^T$
can be expressed as \begin{equation}\label{eq_ce_0}\hat{\mathbf{H}}=\sqrt{1-\rho}\mathbf{Z}_1,\quad\tilde{\mathbf{H}}=\sqrt{\rho}\mathbf{Z}_2,\quad 0\leq\rho\leq 1.\end{equation}$\hat{\mathbf{h}}_k\in\mathbb{C}^{1\times M}$
and $\tilde{\mathbf{h}}_k\in\mathbb{C}^{1\times M}$ are the $k$-th
rows of $\hat{\mathbf{H}}$ and $\tilde{\mathbf{H}}$, respectively.
Both $\mathbf{Z}_1\in\mathbb{C}^{K\times M}$ and $\mathbf{Z}_2\in\mathbb{C}^{K\times M}$
are composed of i.i.d. $\mathcal{CN}(0,1)$ entries, and the two matrices
are independent with each other\footnote{Note that the imperfect CSI model invoked here is similar to that of \cite{yang_performance_2013} and different from that employed by \cite{wanger_large_2012} and \cite{couillet_random_2011}. In \cite{wanger_large_2012} and \cite{couillet_random_2011}, the authors assumed $\hat{\mathbf{H}}=\mathbf{H}+\tilde{\mathbf{H}}$, where $\tilde{\mathbf{H}}$ and $\mathbf{H}$ are mutually independent.}.
We assume that the LSF CSI of each UE and the transmit correlation
matrix $\mathbf{R}$ are perfectly known at the BS. According to \eqref{eq_h_0}
and \eqref{eq_ce_0}, the downlink channel matrix defined in \eqref{eq_g_0}
can be rewritten as\footnote{In fact, as shown in \cite{biguesh_training-based_2006}, the explicit forms of  $\hat{\mathbf{G}}$ and $\tilde{\mathbf{G}}$ is related to the  channel estimation approach, the power and the length of the training pilots, as well as the statistical information of the estimated channel. Here, we adopt a simplified model for tractability. Some tailored user selection schemes relying on a specific channel estimation algorithm (e.g., LS, MMSE) will be postponed for our future work.}\begin{equation}\begin{split}
\mathbf{G}&=\hat{\mathbf{G}}+\tilde{\mathbf{G}}\\
&=\mathbf{D}^{1/2}\hat{\mathbf{H}}\mathbf{R}^{1/2}+\mathbf{D}^{1/2}\tilde{\mathbf{H}}\mathbf{R}^{1/2}.
\end{split}\end{equation}Upon invoking the estimated channel matrix $\hat{\mathbf{G}}$ and
the ZF precoding, the transmitted vector $\mathbf{x}\in\mathbb{C}^{M\times 1}$
is written as \begin{equation}\label{eq_zf_0}\begin{split}
\mathbf{x}&=\gamma\hat{\mathbf{G}}^{\dagger}\mathbf{s}\\
&=\gamma\hat{\mathbf{G}}^H(\hat{\mathbf{G}}\hat{\mathbf{G}}^H)^{-1}\mathbf{s},
\end{split}\end{equation}where $\gamma$ is the power controlling factor, $\mathbf{s}=[s_1,\cdots,s_K]^T$
is the $K\times 1$ information-bearing symbol vector and $s_k$ represents
the symbol intended to the $k$-th active-UE. Denoting the available
total transmit power at the BS as $P$, the long-term power constraint
is given by $$\text{E}[\text{tr}(\mathbf{x}\mathbf{x}^H)]\leq P.$$
Thus, $\gamma$ can be calculated as\begin{equation}\label{eq_pc_0}\gamma=\sqrt{\frac{P}{\text{tr}(\hat{\mathbf{G}}\hat{\mathbf{G}}^H)^{-1}}}.\end{equation}It
should be noted that we employ the equal power allocation among active-UEs
for the sake of low computational complexity. Relying on \eqref{eq_zf_0},
the $K\times 1$ dimensional received signal vector at the $K$ active-UEs
is expressed as \begin{equation}\begin{split}\mathbf{y}&=\mathbf{G}\mathbf{x}+\mathbf{n}\\&=\gamma\mathbf{s}+\gamma\tilde{\mathbf{G}}\hat{\mathbf{G}}^{\dagger}\mathbf{s}+\mathbf{n},\end{split}\end{equation}where
$\mathbf{n}=[n_1,\cdots,n_k]^T\in\mathbb{C}^{K\times 1}$ is the additive
white Gaussian noise (AWGN) vector at the UEs, and $n_k\sim\mathcal{CN}(0,\sigma_n^2)$
represents the noise at the $k$-th active-UE. Additionally, the received
signal received at the $k$-th active-UE is given by \begin{equation}y_k=\gamma s_k+\gamma\sqrt{cd_k^{-\alpha}}\tilde{\mathbf{h}}_k\mathbf{R}^{1/2}\hat{\mathbf{G}}^{\dagger}\mathbf{s}+n_k.\end{equation}Then,
we can write the SINR recorded at the $k$-th active-UE as \begin{equation}\label{eq_sinr_0}\text{SINR}_k=\frac{\gamma^2}{\sigma_n^2+cd_k^{-\alpha}\gamma^2\tilde{\mathbf{h}}_k\mathbf{R}^{1/2}\hat{\mathbf{G}}^{\dagger}\mathbf{s}\mathbf{s}^H(\hat{\mathbf{G}}^{\dagger})^H(\mathbf{R}^{1/2})^H\tilde{\mathbf{h}}_k^H}.\end{equation}Then,
the sum rate $\mathcal{R}$ is given by \begin{equation}\label{eq_r_0}\begin{split}
\mathcal{R}&=\left(1-\frac{K}{T}\right)\sum_{k=1}^K\mathcal{R}_k\\
&=\left(1-\frac{K}{T}\right)\sum_{k=1}^K\log_2(1+\text{SINR}_k),
\end{split}\end{equation}where $\mathcal{R}_k$ is the rate of the $k$-th active-UE, and $T$
denotes the number of symbols over which the channel is constant.
As a percentage, the pre-log factor $(1-K/T)$ implies that the downlink
data transmission only occupies a fraction of the coherence slot.
In particular, we assume that $K$ UEs are simultaneously served.
Since each of the $K$ UEs is assigned one of the $K$ orthogonal
pilot sequences, the length of the pilot sequence should not be shorter
than $K$ symbols\cite{rusek_scaling_2013}. For simplicity, in this
paper we adopt the shortest available pilot sequence of $K$-symbol
length. Because the BS does not transmit data during the uplink pilot
transmission for channel estimation, there exist $(T-K)$ symbols
left for downlink data transmission at each coherence slot consisting
of $T$ symbols. Therefore, the sum rate for downlink data transmission
may be evaluated using \eqref{eq_r_0}. Note that in order to guarantee
the feasibility of data transmission, we assume $K<T$ in this paper.
Otherwise, the pilot transmission would occupy the entire coherence
slot $T$.

\section{Asymptotic Sum Rate Analyses-based Low-Complexity User Selection}

In this section, the proposed user selection schemes are presented.
First, relying on the RMT-based large system analysis, we derive a
deterministic approximation of the ergodic sum rate of the ZF precoder
aided massive MIMO system. This result brings new insights into the
question of how to enhance the system sum rate performance. Then,
a pair of low-complexity user selection schemes are proposed based
on the attained approximation of the sum rate.

\subsection{Sum Rate Approximation in the Large-System Regime}

We first evaluate the value of $\text{SINR}_k$ for the scenario where
$M\text{ and }K$ go to infinity with a finite ratio $M/K>0$. Then,
the approximation of the ergodic sum rate is derived.

Applying RMT-based large-system analysis, we reveal that $\text{SINR}_k$
may be characterized by (please see Appendix for the detailed derivation)
\begin{equation}\label{eq_sinr_1}
\text{SINR}_k\xrightarrow{a.s.}\frac{1}{\sum_{i=1}^Kd_i^{\alpha}\left(A(K,M)+B(K,M)d_k^{-\alpha}\right)},
\end{equation}where \begin{equation}\label{eq_ab_0}\begin{split}
A(K,M)&=\frac{1}{1-\rho}\frac{\sigma_n^2}{Pc\phi M},\\
B(K,M)&=\frac{\rho}{1-\rho}\frac{\psi}{M\phi^2-K\psi}.
\end{split}\end{equation}In \eqref{eq_ab_0}, $\phi$ is the unique solution of the equation
$$
\phi=\frac{1}{M}\text{tr}\left(\mathbf{R}\left(\mathbf{I}_M+\frac{K}{M}\frac{1}{\phi}\mathbf{R}\right)^{-1}\right)
$$and $\psi$ is defined as $$\psi=\frac{1}{M}\text{tr}\left(\mathbf{R}^2\left(\mathbf{I}_M+\frac{K}{M}\frac{1}{\phi}\mathbf{R}\right)^{-2}\right).$$

Exploiting \eqref{eq_r_0} and \eqref{eq_sinr_1}, the ergodic sum
rate $\text{E}[\mathcal{R}]$ in large-system regime can be formulated
as \begin{equation}\label{eq_lb_0}\begin{split}
\text{E}[\mathcal{R}]&=\left(1-\frac{K}{T}\right)\text{E}_{d_1,\cdots,d_K}\left[\sum_{k=1}^K\log_2(1+\text{SINR}_k)\right]\\
&\geq\tilde{\mathcal{R}},
\end{split}\nonumber\end{equation}in which $\tilde{\mathcal{R}}$ is defined as \begin{equation}\label{eq_rsim_0}\tilde{\mathcal{R}}=\left(1-\frac{K}{T}\right)\sum_{k=1}^K\text{E}_{d_k}\left[\log_2(1+\widetilde{\text{SINR}}_k)\right],\end{equation}and
``$\geq$'' is obtained by applying Jensen's inequality$$\text{E}\left[\log_2\left(1+\frac{1}{x}\right)\right]\geq \log_2\left(1+\frac{1}{\text{E}[x]}\right).$$Additionally,
$\widetilde{\text{SINR}}_k$ is given by $$
\widetilde{\text{SINR}}_k=\frac{1}{T_1d_k^{\alpha}+T_2d_k^{-\alpha}+T_3},
$$where \begin{equation}\label{eq_t123_0}\begin{split}
T_1&=A(K,M),\\
T_2&=B(K,M)\sum_{i=1,i\neq k}^K\text{E}_{d_i}[d_i^{\alpha}],\\
T_3&=A(K,M)\sum_{i=1,i\neq k}^K\text{E}_{d_i}[d_i^{\alpha}]+B(K,M).
\end{split}\end{equation}

In this paper, we employ the lower bound $\tilde{\mathcal{R}}$ defined
by \eqref{eq_lb_0} as an approximation of the system sum rate. It
should be emphasized that $\tilde{\mathcal{R}}$ is independent of
the SSF CSI of UEs, which constitute the basis for our designs.
\begin{rem}
The results obtained above are invalid for the scenario of $K=M$
due to mathematical intractability. Fortunately, as shown by our simulation
results that are given in Section \ref{sub:Sum-Rate-Performance},
$K=M$ is rarely beneficial for enhancing the performance of the considered
massive MIMO system. This phenomenon was also observed by\cite{bjornson_optimal_2014}
and \cite{wanger_large_2012}. Therefore, in this section we focus
on the sum rate approximation for $K<M$.
\end{rem}

\subsection{$K^*$-Based Random User Selection ($K^*$-RUS)\label{sub:Random--User}}

In this subsection, we develop a novel RUS scheme, namely $K^*$-RUS,
for the sake of improving the system sum rate and ensuring the fairness
among candidate-UEs. Specifically, compared with the conventional
RUS scheme in which $M$ UEs are selected for simultaneous data transmissions
at each coherence slot, we modify the number of active-UEs to a more
appropriate value $K^*$, which is decided according to the system
parameters (e.g., the transmit power $P$ of the BS) and the statistical
information of the channel (e.g., the probability distributions of
SSF CSI and LSF CSI).
\begin{figure}[tbh]
\begin{centering}
\includegraphics[width=5cm]{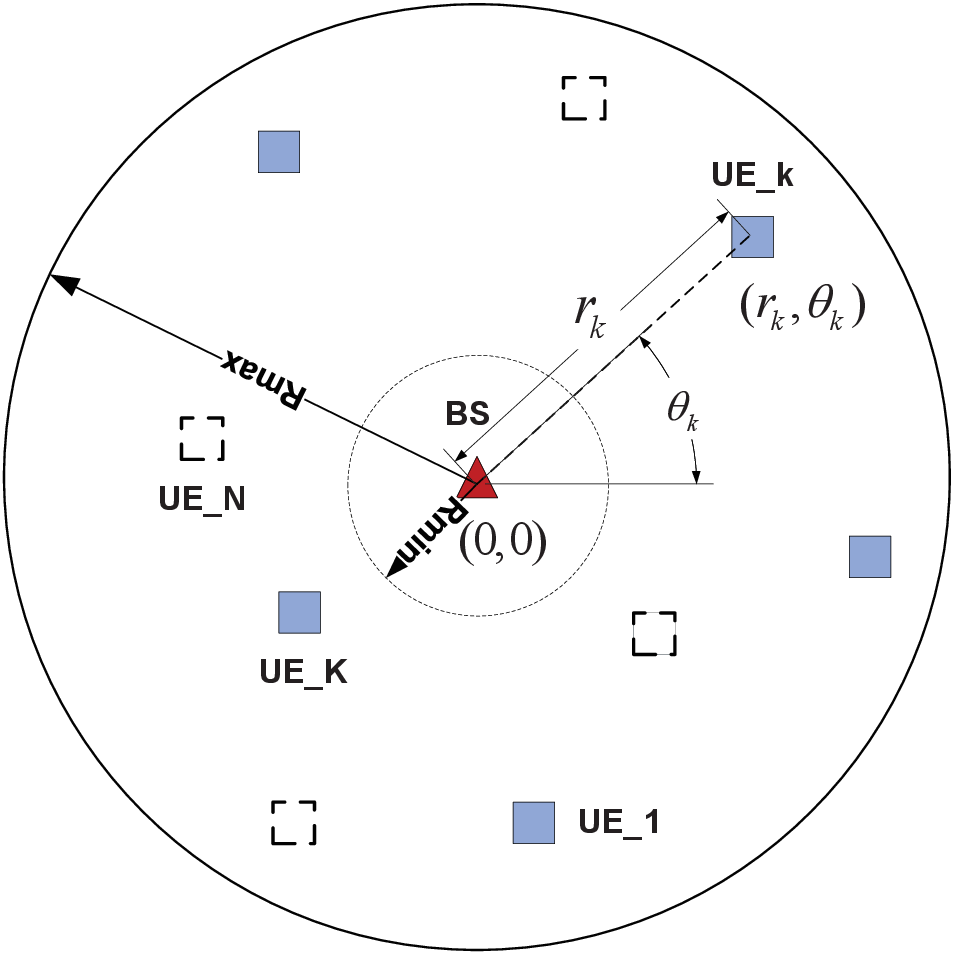}
\par\end{centering}

\protect\caption{The diagram of UE locations.}
\label{fig_ue}
\end{figure}

In order to obtain $K^*_{\text{RUS}}$, we need to consider the user
distribution, which facilitates characterizing the effects of LSF.
As shown in Fig. \ref{fig_ue}, in this paper we employ the common
circular cell model \cite{baltzis_hexagonal_2011}, where all the
$N$ candidate-UEs are independently uniformly distributed (i.u.d.)
in a circular cell having an inner radius of $R_{\min}$ and outer
radius of $R_{\max}$, while the BS is located at the center point.
We denote the locations in polar coordinates, i.e., $(r_k,\theta_k)$
is the location for the $k$-th candidate-UE and $(0,0)$ for the
BS. Therefore, the probability density functions (PDFs) of $r_k$
and $\theta_k$ are given by (the subscript $k$ is omitted below
for ease of notation) \begin{equation}\label{eq_df_0}f_R(r)=\frac{2r}{R_{\max}^2-R_{\min}^2},\quad R_{\min}\leq r\leq R_{\max},\end{equation}
and \begin{equation}f_{\Theta}(\theta)=\frac{1}{2\pi},\nonumber\end{equation}respectively.
The cumulative distribution functions (CDFs) of $r_k$ and $\theta_k$
are written as \begin{equation}\label{eq_df_1}F_R(r)=\frac{r^2-R_{\min}^2}{R_{\max}^2-R_{\min}^2},\quad R_{\min}\leq r\leq R_{\max},\end{equation}
and \begin{equation}F_{\Theta}(\theta)=\frac{\theta}{2\pi},\nonumber\end{equation}respectively.

As a result, for $K^*$-RUS, the PDF of the distance from the $k$-th
active-UE to the BS, i.e. $f_{d_{k}}^{\text{RUS}}(r)$, is given by
$$f_{d_{k}}^{\text{RUS}}(r)=f_R(r),\quad R_{\min}\leq r\leq R_{\max},\quad k=1,\dots,K.$$Then,
it is easy to obtain \begin{equation}\label{eq_t123_1}\begin{split}
\sum_{i=1,i\neq k}^K\text{E}_{d_i}[d_i^{\alpha}]&=(K-1)\int_{R_{\min}}^{R_{\max}}r^{\alpha}f_R(r)dr\\
&=\frac{2(K-1)(R_{\max}^{\alpha+2}-R_{\min}^{\alpha+2})}{(\alpha+2)(R_{\max}^2-R_{\min}^2)}.
\end{split}\end{equation}Thus, $\tilde{\mathcal{R}}$ in \eqref{eq_rsim_0} can be rewritten
as \begin{equation}\label{eq_rus_0}\begin{split}
\tilde{\mathcal{R}}_{\text{RUS}}&=\left(1-\frac{K}{T}\right)K\\
&\times\int_{R_{\min}}^{R_{\max}}\log_2\left(1+\frac{1}{T_1r^{\alpha}+T_2r^{-\alpha}+T_3}\right)f_{d_k}^{\text{RUS}}(r)dr.
\end{split}\end{equation}Substituting \eqref{eq_ab_0} and \eqref{eq_t123_1} into \eqref{eq_t123_0},
and then substituting \eqref{eq_t123_0} and \eqref{eq_df_0} into
\eqref{eq_rus_0}, $\tilde{\mathcal{R}}_{\text{RUS}}$ can be reformulated
as \eqref{eq_rus_final} which is given on the next page.
\begin{figure*}[tbh]
\begin{equation}\label{eq_rus_final}\begin{split}
&\tilde{\mathcal{R}}_{\text{RUS}}\\
&=\Gamma_1(T,M,K,R_{\min},R_{\max},P,c,\sigma_n^2,\rho,\alpha,\delta)\\
&=\left(1-\frac{K}{T}\right)K\\
&\times\int_{R_{\min}}^{R_{\max}}\frac{2r}{R_{\max}^2-R_{\min}^2}
\log_2\left(1+\frac{1-\rho}{\frac{\sigma_n^2}{Pc\phi M}\left(r^{\alpha}+\frac{2(K-1)(R_{\max}^{\alpha+2}-R_{\min}^{\alpha+2})}{(\alpha+2)(R_{\max}^2-R_{\min}^2)}\right)+\frac{\rho\psi}{M\phi^2-K\psi}\left(\frac{2(K-1)(R_{\max}^{\alpha+2}-R_{\min}^{\alpha+2})}{(\alpha+2)(R_{\max}^2-R_{\min}^2)}r^{-\alpha}+1\right)}\right)dr.
\end{split}\end{equation}

\rule[0.5ex]{2\columnwidth}{0.5pt}
\end{figure*}
 We can see that \eqref{eq_rus_final} is only related to the system
parameters of $T,M,K,R_{\min},R_{\max},P,c,\sigma_n^2,\rho,\alpha\text{ and }\delta$.

Given these system parameters, the optimal number $K^*_{\text{RUS}}$
in the sense of $\tilde{\mathcal{R}}_{\text{RUS}}$ maximization can
be obtained efficiently with an one-dimensional search over the candidate
set $\{1,2,\dots,M-1\}$, i.e., \begin{equation}\label{eq_krus_0}\begin{split}
&K^*_{\text{RUS}}\\
&=\mathop{\arg\max}_{K\in\{1,\dots,M-1\}}\Gamma_1(T,M,K,R_{\min},R_{\max},P,c,\sigma_n^2,\rho,\alpha,\delta).
\end{split}\end{equation}Obviously, $K^*_{\text{RUS}}$ is independent of any instantaneous
CSI, which makes it possible to find $K^*_{\text{RUS}}$ offline.

After obtaining $K^*_{\text{RUS}}$, the only online operation in
the proposed $K^*$-RUS scheme is to randomly select $K^*_{\text{RUS}}$
UEs for simultaneous data transmissions. Therefore, little extra computational
complexity is imposed on the proposed $K^*$-RUS compared to the conventional
RUS.
\begin{rem}
The authors of \cite{wanger_large_2012,zhao_performance_2013} and
\cite{jung_optimal_2013} also discussed the optimal number of UEs
for the ZF precoding, but they treated the LSF coefficients of UEs
as deterministic values, which limit the generality of $K^*$. In
other words, $K^*$ has to be updated whenever any LSF CSI of the
system changes. By contrast, we take random UE locations into account
for obtaining a more general and practical $K^*$. As long as the
statistical properties of the system remain unchanged, our $K^*$
keeps its current value for arbitrary LSF and SSF channel realizations.
\end{rem}

\subsection{$K^*$-Based Location-Dependant User Selection ($K^*$-LUS)\label{sub:-Based-Location-Dependant-User}}

In order to enhance the sum rate performance further, a $K^*$-based
location-dependant user selection scheme, namely the $K^*$-LUS is
developed in this subsection. In $K^*$-LUS, we select an appropriate
number of UEs in descending order of the LSF coefficients (i.e., in
ascending order of the BS-UE distances) for simultaneous data transmissions.

Inspired by $K^*$-RUS, let us first investigate the optimal number
of active-UEs in LUS, which is denoted by $K^*_{\text{LUS}}$. We
assume that $K$ UEs are selected and the distances between the selected
UEs and the BS satisfy $d_1\leq d_2\leq\cdots\leq d_K$. Let $(r_k,\theta_k)$
represent the $k$-th active-UE according to the distance in ascending
order. According to \eqref{eq_df_1}, we can get the PDF of the order
statistic $d_k,k=1,\dots,K$ in LUS as \begin{equation}\label{eq_df_3}\begin{split}
f_{d_k}^{\text{LUS}}\!(r_k)\!
&\!=\!\frac{1}{\text{B}(k,N\!-\!k+1)}F_R^{(k-1)}(r_k)[1\!-\!F_R(r_k)]^{N-k}f_R(r_k)\\
&\!=\!\frac{2r_k(R_{\max}^2-r_k^2)^{N-k}(r_k^2-R_{\min}^2)^{k-1}}{\text{B}(k,N-k+1)(R_{\max}^2-R_{\min}^2)^N},
\end{split}\end{equation}in which $\text{B}(x,y)$ represents the Beta function with parameters
$x\text{ and }y$. Then, we have \begin{equation}\label{eq_t123_4}
\text{E}_{d_k}[d_k^{\alpha}]=R_{\min}^{\alpha}{}_2F_1\left(k,-\frac{\alpha}{2};N+1;1-\frac{R_{\max}^2}{R_{\min}^2}\right)\end{equation}and \begin{equation}\label{eq_rlus_0}\begin{split}
&\text{E}_{d_k}\left[\log_2(1+\widetilde{\text{SINR}}_k)\right]\\
&=\int_{R_{\min}}^{R_{\max}}\log_2\left(1+\frac{1}{T_1r_k^{\alpha}+T_2r_k^{-\alpha}+T_3}\right)f_{d_k}^{\text{LUS}}(r_k)dr_k,
\end{split}\end{equation}where ${}_2F_1(\cdot)$ is the ordinary hypergeometric function\cite{gauss_F12}.
Therefore, $\tilde{\mathcal{R}}_{\text{LUS}}$ can be formulated with
the aid of \eqref{eq_ab_0}, \eqref{eq_t123_0}, \eqref{eq_df_3},
\eqref{eq_t123_4} and \eqref{eq_rlus_0}, as shown in \eqref{eq_lus_final}
on the next page.
\begin{figure*}[tbh]
\begin{equation}\label{eq_lus_final}\begin{split}
&\tilde{\mathcal{R}}_{\text{LUS}}\\
&=\Gamma_2(T,M,K,R_{\min},R_{\max},P,c,\sigma_n^2,\rho,\alpha,\delta)\\
&=\left(1-\frac{K}{T}\right)\sum_{k=1}^K
\int_{R_{\min}}^{R_{\max}}\frac{2r_k(R_{\max}^2-r_k^2)^{N-k}(r_k^2-R_{\min}^2)^{k-1}}{\text{B}(k,N-k+1)(R_{\max}^2-R_{\min}^2)^N}
\log_2\Bigg(1+\\
&\quad\frac{1-\rho}{\frac{\sigma_n^2}{Pc\phi M}\left(r_k^{\alpha}+R_{\min}^{\alpha}{}_2F_1\left(k,-\frac{\alpha}{2};N+1;1-\frac{R_{\max}^2}{R_{\min}^2}\right)\right)+\frac{\rho\psi}{M\phi^2-K\psi}\left(r_k^{-\alpha}R_{\min}^{\alpha}{}_2F_1\left(k,-\frac{\alpha}{2};N+1;1-\frac{R_{\max}^2}{R_{\min}^2}\right)+1\right)}\Bigg)dr_k.
\end{split}\end{equation}

\rule[0.5ex]{2\columnwidth}{0.5pt}
\end{figure*}

If the system parameters are fixed, we are capable of solving for
the optimal number $K^*_{\text{LUS}}$ maximizing $\tilde{\mathcal{R}}_{\text{LUS}}$
offline relying on standard line search algorithms, i.e.,\begin{equation}\label{eq_lus_0}\begin{split}
&K^*_{\text{LUS}}\\
&=\mathop{\arg\max}_{K\in\{1,\dots,M-1\}}\Gamma_2(T,M,K,R_{\min},R_{\max},P,c,\sigma_n^2,\rho,\alpha,\delta).
\end{split}\end{equation}As long as we find $K^*_{\text{LUS}}$, we just need to sort UEs in
ascending order of  their distances from the BS and select the first
$K^*_{\text{LUS}}$ UEs for data transmission.
\begin{rem}
With respect to the number of active-UEs $K$, we determine $K=K^*_{\text{RUS}}$
and $K=K^*_{\text{LUS}}$ according to the given system parameters
in the proposed $K^*$-RUS and $K^*$-LUS schemes, respectively. By
contrast, there usually exists the scenario of $K=M$, where the full-spatial-multiplexing
transmission may be performed in the conventional RUS and LUS schemes.
In SUS, the value of $K^*$ depends on specific channel realizations
and will not be obtained until the selection procedure is completed
(please see Section IV-B and Section VI-A in \cite{yoo_optimality_2006}
for more details).
\end{rem}

\subsection{Computational Complexity Analysis }

For a system having $M$ BS antennas and $N$ candidate-UEs, although
the user selection relying on exhaustive search achieves the best
sum rate performance, approximately $\sum_{k=1}^{M}\tbinom{N}{k}k^5M$
complex-valued operations are required to complete one selection\cite{sjobergh_greedy_2008},
which may be unaffordable in practice. For SUS, the computational
complexity is roughly $\mathcal{O}(M^3N)$ \cite{yoo_optimality_2006},
which is high for large-$M$ systems. In stark contrast to these conventional
schemes, the online computational complexity of the proposed $K^*$-RUS
and $K^*$-LUS is independent of $M$ and $N$, and the instantaneous
CSI-based complicated online computations are avoided. As a result,
the computational complexity is approximately $\mathcal{O}(1)$, which
is just the same as that of the conventional RUS scheme.

\subsection{Performance Analysis for the Special Case of $\rho=0,\delta=0$\label{sub:Performance-Analysis}}

All the above investigations are subject to the general case, i.e.,
in the context of the systems with imperfect CSI and transmit antenna
correlation. In this subsection, we consider a special case in which
there exists neither channel estimation error nor transmit antenna
correlation, i.e., $\rho=0,\delta=0$. In this context, because we
can obtain clearer insights into how the system performance is affected
by different user selection schemes.

For $\rho=0,\delta=0$, we have \begin{equation}\begin{split}
A(K,M)&=\frac{\sigma_n^2}{Pc(M-K)},\\
B(K,M)&=0.
\nonumber\end{split}\end{equation}Substituting them into \eqref{eq_rus_final} and \eqref{eq_lus_final},
the approximate sum rates can be calculated. Nevertheless, the integrals
of logarithmic functions in \eqref{eq_rus_final} and \eqref{eq_lus_final}
degrade the intelligibility of the results.

Here, a new method, which is different from those adopted in Section
\ref{sub:Random--User} and Section \ref{sub:-Based-Location-Dependant-User},
is developed for finding a much simpler expression of the ergodic
sum rate approximation in this special case. In what follows we apply
Jensen's inequality in a slightly different manner for the sake of
finding a more concise expression of the system sum rate approximation.
More specifically, we have \begin{equation}\begin{split}
\text{E}[\mathcal{R}]&=\left(1-\frac{K}{T}\right)\text{E}_{d_1,\dots,d_K}\left[\sum_{k=1}^{K}\log_2(1+\text{SINR}_k)\right]\\
&\geq\left(1-\frac{K}{T}\right)K\log_2(1+\widetilde{\widetilde{\text{SINR}}}),
\end{split}\nonumber\end{equation}where $$\widetilde{\widetilde{\text{SINR}}}=\frac{Pc(M-K)}{\sigma_n^2\sum_{k=1}^K\text{E}_{d_k}[d_k^{\alpha}]}.$$The
approximate sum rate $\tilde{\tilde{\mathcal{R}}}$ for $K^*$-RUS
is then given by \begin{equation}\label{eq_s4_1}\begin{split}
&\tilde{\tilde{\mathcal{R}}}^*_{\text{RUS}}=\left(1-\frac{K^*_{\text{RUS}}}{T}\right)K^*_{\text{RUS}}\\
&\times\log_2\left(1+\frac{Pc(M-K^*_{\text{RUS}})(\alpha+2)(R_{\max}^2-R_{\min}^2)}{2\sigma_n^2K(R_{\max}^{\alpha+2}-R_{\min}^{\alpha+2})}\right).
\end{split}\end{equation}For $K^*$-LUS, the sum rate is approximated as \begin{equation}\label{eq_s4_2}\begin{split}&\tilde{\tilde{\mathcal{R}}}^*_{\text{LUS}}=\left(1-\frac{K^*_{\text{LUS}}}{T}\right)K^*_{\text{LUS}}\\&\times\log_2\!\!\left(\!\!1\!+\!\frac{Pc(M-K^*_{\text{LUS}})}{\sigma_n^2\sum_{k=1}^{K^*_{\text{LUS}}}R_{\min}^{\alpha}{}_2F_1\left(k,-\frac{\alpha}{2};N\!\!+\!\!1;1-\frac{R_{\max}^2}{R_{\min}^2}\!\!\right)}\!\!\right).\end{split}\end{equation}Compared
to \eqref{eq_rus_final} and \eqref{eq_lus_final}, there is no integral
calculation of logarithmic functions in \eqref{eq_s4_1} and \eqref{eq_s4_2},
which simplifies the system sum rate expression. Furthermore, as shown
in Fig. \ref{fig3}, \eqref{eq_s4_1} and \eqref{eq_s4_2} are capable
of providing tight approximations in the case of $\rho=0,\delta=0$.

According to \eqref{eq_s4_1} and \eqref{eq_s4_2}, it is clear that
both of the sum rates $\tilde{\tilde{\mathcal{R}}}^*_{\text{RUS}}$
and $\tilde{\tilde{\mathcal{R}}}^*_{\text{LUS}}$ increase when $P$
and $M$ become larger. Additionally, when $N$ increases, the sum
rate of $K^*$-LUS increases, while the sum rate of $K^*$-RUS remains
unchanged.

As far as SUS is concerned in the TDD scenario, as proved in \cite{yoo_optimality_2006},
the ergodic sum rate is upper bounded by $\bar{\mathcal{R}}_{\text{SUS}}$,
which is given by \begin{equation}\label{eq_s4_3}\bar{\mathcal{R}}_{\text{SUS}}=\left[1-\frac{N}{T}\right]^+M\Theta(\log_2\log_2N),\end{equation}
where $[\cdot]^+$ is defined as $[x]^+=\max\{x,0\}$. Due to the
pilot overhead imposed by the channel estimation, the system sum rate
is scaled by the factor $[1-N/T]^+$. More specifically, in this case
the instantaneous CSI of all the $N$ candidate-UEs are required for
select ingactive-UEs at each coherence slot.

Comparing the pre-log parts $[1-N/T]^+$ in \eqref{eq_s4_3}, $(1-K^*_{\text{RUS}}/T)$
in \eqref{eq_s4_1} and $(1-K^*_{\text{LUS}}/T)$ in \eqref{eq_s4_2},
we can see that the advantages of $K^*$-RUS and $K^*$-LUS are obvious
for the systems relying on pilot-based channel estimation. In particular,
when $T$ is not significantly larger than $N$, with regard to SUS,
a large portion of the coherence slot would be dedicated to channel
estimation, which reduces the resources for the downlink data delivery.
By contrast, with the proposed user selection schemes, we only have
to estimate the CSI of the $K^*_{\text{RUS}}$ or $K^*_{\text{LUS}}$
active-UEs for precoding. Both $K^*_{\text{RUS}}$ and $K^*_{\text{LUS}}$
are usually much smaller than $N$, hence our schemes are superior
to SUS by exploiting more data transmission resources.
\begin{rem}
The result of $\Theta(\log_2\log_2N)$ in \eqref{eq_s4_3} is obtained
when we have $N\to\infty$ and a fixed $M$\cite{yoo_optimality_2006}.
Moreover, as shown in \cite{tomasoni_selection_2009}, when the number
of candidate-UEs $N$ is linearly related to the number of BS antennas
$M$ (i.e. the case we have discussed in this paper), only marginal
multiuser diversity gain might be achieved by SUS.
\end{rem}

\section{Numerical Simulations and Discussions \label{sec:Numerical-Simulations}}

In this section we present simulation results to show the benefits
of the proposed user selection schemes. The cellular model employed
is based on that of \cite{bjornson_optimal_2014} and \cite{3GPP_1}.
Like \cite{yang_performance_2013}, we assume the number of symbols
in a coherence slot is $196$, over which the channel is constant,
i.e., $T=196$. The cell radius is $R_{\max}=250$ m and the minimum
distance is $R_{\min}=35$ m. The pathloss exponent is $\alpha=3.76$
and the reference LSF factor is $c=10^{-3.53}$. The total noise power
is assumed as $\sigma_{\text{n}}^2=-96$ dBm.

\subsection{Sum Rate Performance\label{sub:Sum-Rate-Performance}}

\begin{figure}[t]
\begin{centering}
\includegraphics[width=7.5cm]{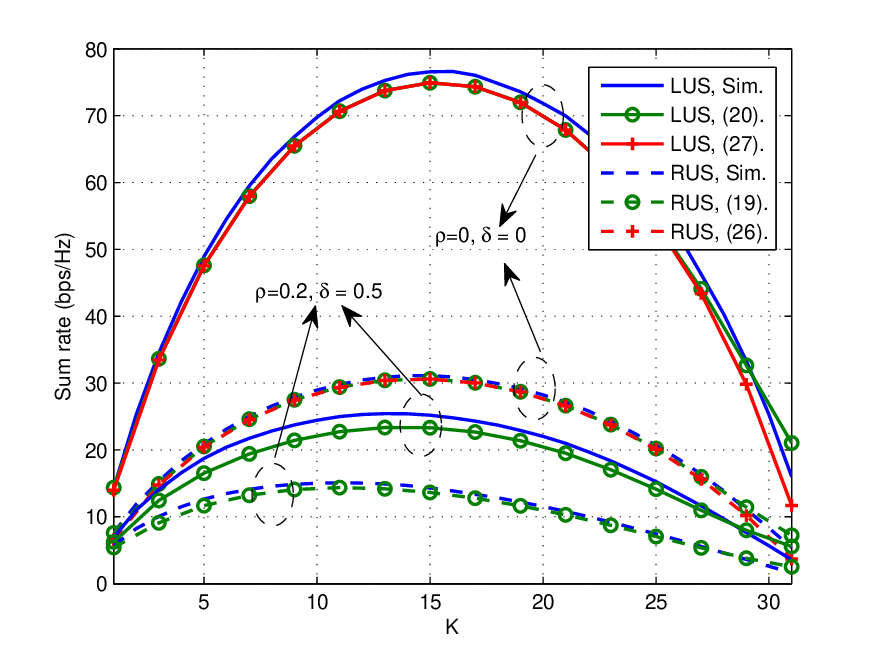}
\par\end{centering}

\centering{}\protect\caption{Sum rates of $K$-RUS and $K$-LUS. $P=30\text{ dBm},M=32,N=64.$}
\label{fig3}
\end{figure}

\begin{figure}[t]
\begin{centering}
\includegraphics[width=7.5cm]{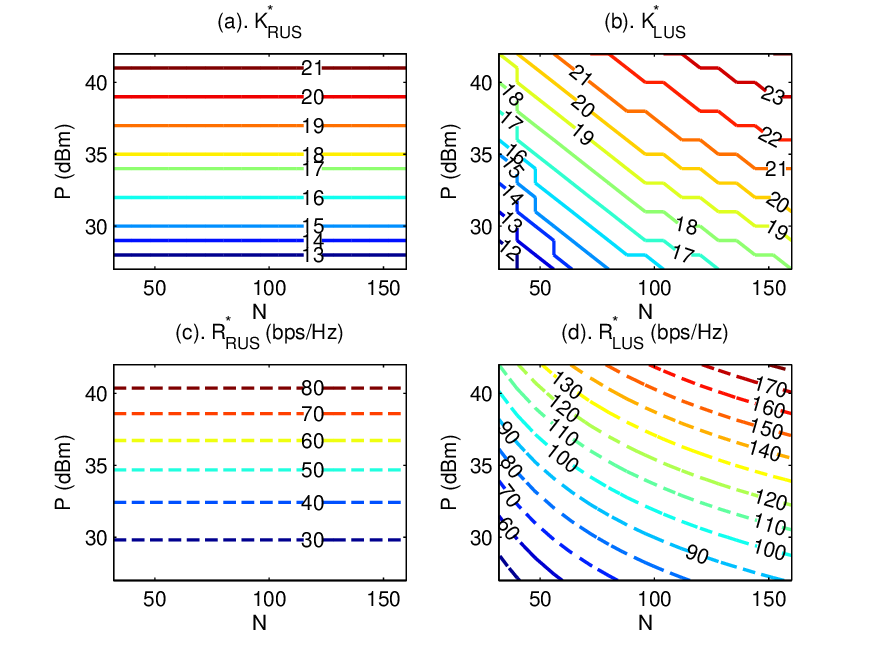}
\par\end{centering}

\centering{}\protect\caption{Contour figures of $K^*_{\text{RUS}}$, $K^*_{\text{LUS}}$, $R^*_{\text{RUS}}$
and $R^*_{\text{LUS}}$ against $P\text{ and }N$. $\rho=0,\delta=0,M=32.$
$K^*_{\text{RUS}}$ and $K^*_{\text{LUS}}$ are shown in (a) (i.e.,
the top-left sub figure) and (b) (i.e., the top-right sub figure)
with solid lines, respectively. $R^*_{\text{RUS}}$ and $R^*_{\text{LUS}}$
are shown in (c) (i.e., the bottom-left sub figure) and (d) (i.e.,
the bottom-right sub figure) with dash lines, respectively. }
\label{fig4}
\end{figure}

\begin{figure}[t]
\begin{centering}
\includegraphics[width=7.5cm]{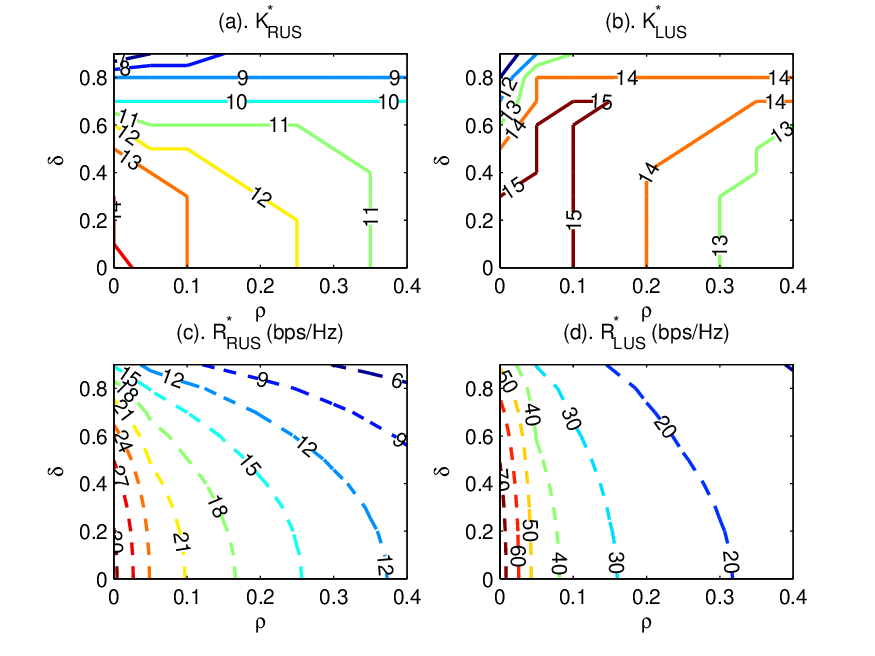}
\par\end{centering}

\centering{}\protect\caption{Contour figures of $K^*_{\text{RUS}},K^*_{\text{LUS}},R^*_{\text{RUS}}$
and $R^*_{\text{LUS}}$ against $\rho\text{ and }\delta$. $P=30\text{ dBm},M=32,N=64.$
$K^*_{\text{RUS}}$ and $K^*_{\text{LUS}}$ are shown in (a) and (b)
with solid lines, respectively. $R^*_{\text{RUS}}$ and $R^*_{\text{LUS}}$
are shown in (c) and (d) with dash lines, respectively. }
\label{fig4-1}
\end{figure}

In Fig. \ref{fig3}, the sum rate performance of $K$-RUS and $K$-LUS
as a function of $K$ is evaluated in the case of $P=30$ dBm, $M=32$
and $N=64$. The simulated ergodic sum rates (marked as 'Sim.' in
the figure) are obtained by averaging over 10000 independent channel
realizations (both SSF and LSF CSI are regenerated at each realization).
It is observed that there exist $K^*_{\text{RUS}}$ and $K^*_{\text{LUS}}$
which maximize the system sum rate for $K$-RUS and $K$-LUS, respectively.
Moreover, we have $K^*_{\text{RUS}}<M$ and $K^*_{\text{LUS}}<M$
for the simulation parameters considered. In Fig. \ref{fig3} we also
show the approximate sum rates given by \eqref{eq_rus_final} and
\eqref{eq_lus_final}, which are very tight. Therefore, it is reasonable
to design user selection schemes based on them.

In order to provide intuitive insights into how $K^*_{\text{RUS}}$
and $K^*_{\text{LUS}}$ are affected by $P$ and $N$, we evaluated
$K^*_{\text{RUS}}$ and $K^*_{\text{LUS}}$ with no transmit correlation
and perfect CSI estimation (i.e., $\rho=0, \delta=0$) in Fig. \ref{fig4}
(a) and (b), respectively. It is observed that both $K^*_{\text{RUS}}$
and $K^*_{\text{LUS}}$ increase with the transmit power $P$. On
the other hand, when we increase $N$, $K^*_{\text{LUS}}$ increases
and $K^*_{\text{RUS}}$ remains unchanged. This can be easily explained
from the perspective of multiuser diversity gain. In particular, $K^*_{\text{LUS}}$
is related to $N$ because we select UEs according to LSF in $K^*$-LUS.
By contrast, $K^*_{\text{RUS}}$ is independent of $N$ because random
selection is adopted in $K^*$-RUS. Moreover, Fig. \ref{fig4} (c)
and Fig. \ref{fig4} (d) provide the sum rates of $K^*$-RUS and $K^*$-LUS.
As expected, we can see that both $\mathcal{R}^*_{\text{RUS}}$ and
$\mathcal{R}^*_{\text{LUS}}$ are enhanced by the increase of $P$.
When $N$ increases, $\mathcal{R}^*_{\text{LUS}}$ rises but $\mathcal{R}^*_{\text{RUS}}$
remains unchanged.

In Fig. \ref{fig4-1} the impacts of channel estimation accuracy $\rho$
and channel correlation factor $\delta$ on the system performance
are characterized, where we set $P=30$ dBm, $M=32$ and $N=64$.
In sub-figures (a) and (b), the optimal number of active-UEs for $K^*$-RUS
and $K^*$-LUS are shown, respectively. We can see that both $K^*_{\text{RUS}}$
and $K^*_{\text{LUS}}$ attain their maximum values at $\rho=0,\delta=0$.
This observation indicates that the BS should serve more UEs under
uncorrelated channel scenarios with perfect CSI estimation than those
under correlated channel scenarios with imperfect CSI estimation.
The sum rates of $K^*$-RUS and $K^*$-LUS are shown in the sub-figures
(c) and (d), respectively. It is clear that the sum rates decrease
as $\delta$ and $\rho$ increase.

\begin{figure}[t]
\begin{centering}
\includegraphics[width=7.5cm]{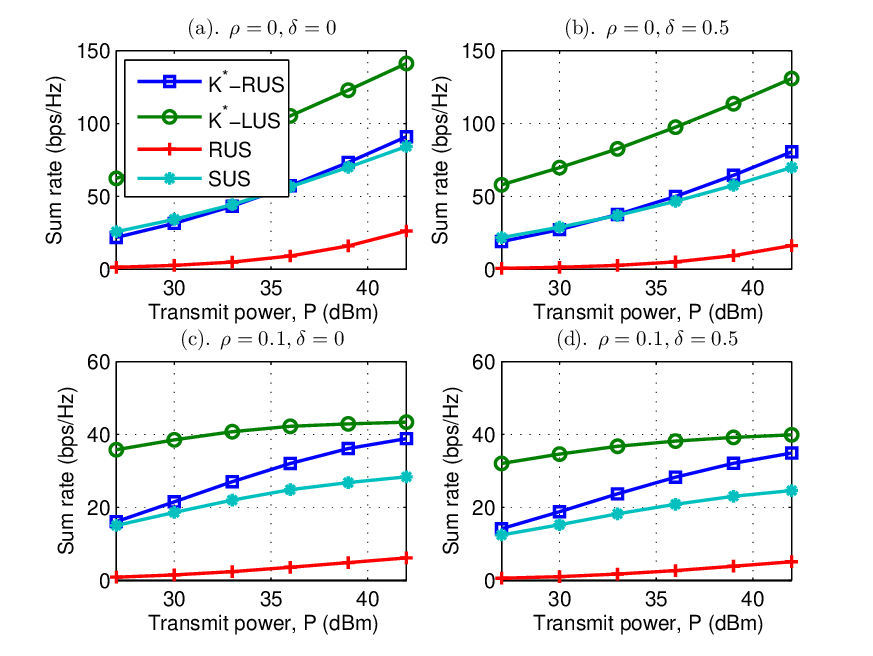}
\par\end{centering}

\centering{}\protect\caption{Sum rate $\mathcal{R}$ vs. transmit power $P$.
$M=32\text{ and }N=64.$}
\label{fig7}
\end{figure}

The sum rate performance of various user selection schemes, including
$K^*$-LUS, $K^*$-RUS, SUS and RUS, is evaluated against the transmit
power $P$ with $M=32,N=64$ in Fig. \ref{fig7}. Four scenarios are
investigated, i.e., uncorrelated channel with perfect CSI estimation
($\rho=0, \delta=0$), correlated channel with perfect CSI estimation
($\rho=0, \delta=0.5$), uncorrelated channel with imperfect CSI estimation
($\rho=0.1, \delta=0$) and correlated channel with imperfect CSI
estimation ($\rho=0.1, \delta=0.5$). Note that for the conventional
RUS, we randomly select $M$ UEs for simultaneous data transmissions.
For SUS, in order to ensure fair comparisons among all the four schemes,
we adopt the equal power allocation instead of the water filling  allocation.
Furthermore, the optimal value of $\alpha_{\text{SUS}}$, which is
an important parameter in SUS (described as $\alpha$ in \cite{yoo_optimality_2006}),
is used for the SUS scheme in our simulations\footnote{The sum rate performance of the SUS scheme is highly sensitive to the choice of  $\alpha_{\text{SUS}}$. The optimal value of $\alpha_{\text{SUS}}$ varies with the changes of $M,N$ and $P$. By means of searching over the interval $(0,1]$, we obtain optimal values of $\alpha_{\text{SUS}}$ maximizing the sum rates of the SUS scheme for different configurations of  $M,N$ and $P$.}.
As expected, $K^*$-LUS achieves the best sum rate performance among
the four schemes, and compared with RUS, $K^*$-RUS also achieves
a significant sum rate improvement. In addition, due to the non-negligible
channel estimation pilot overhead and the lack of multiuser diversity
gain, the conventional SUS scheme achieves similar {[}e.g., in sub-figures
(a) and (b){]} or even worse {[}e.g., in sub-figures (c) and (d){]}
sum rate performance than the proposed $K^*$-RUS, even though the
online computational complexity of SUS is much higher than that of
$K^*$-RUS.

\begin{figure}[t]
\begin{centering}
\includegraphics[width=7.5cm]{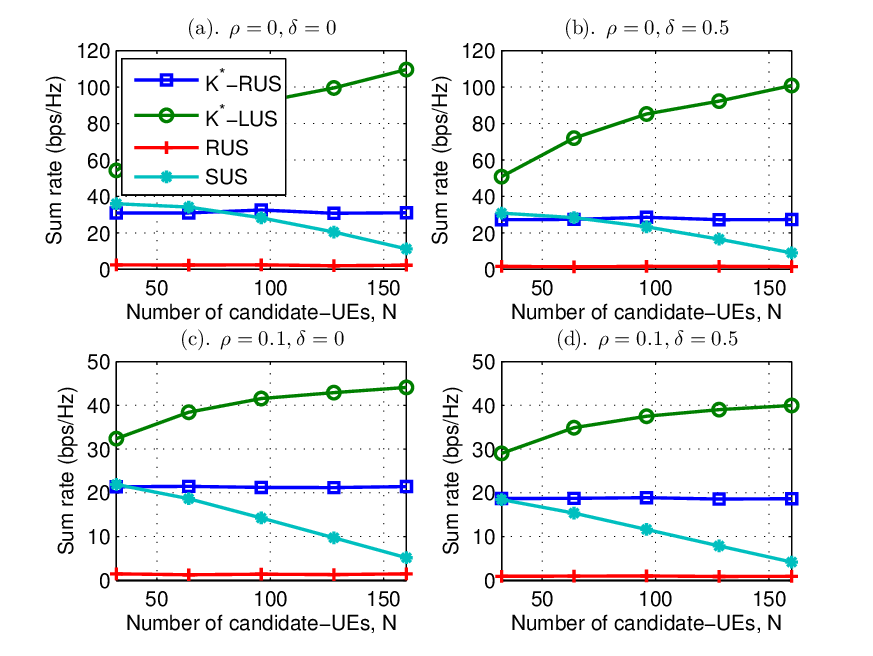}
\par\end{centering}

\centering{}\protect\caption{Sum rate $\mathcal{R}$ vs. the number of candidate-UEs
$N$. $P=30 \text{ dBm and }M=32.$}
\label{fig8}
\end{figure}

Furthermore, in Fig. \ref{fig8}, we investigate the sum rates of
the four schemes for different numbers of candidate-UEs, $N$, when
$P=30$ dBm and $M=32$. It is clear that with the increase of $N$,
the sum rates of $K^*$-RUS and RUS remain unchanged. In contrast,
$K^*$-LUS obtains sum rate improvements as $N$ increases because
more multiuser diversity related to LSF can be exploited with larger
$N$. For SUS, the sum rate decreases with $N$, because the pilot
overhead becomes serious for large $N$, which overwhelms the increase
of the multiuser diversity gain.

\subsection{Fairness Performance}

In this subsection, we evaluate the system performance in terms of
long-term fairness among UEs. The Jain's Fairness Index (JFI)\cite{DEC_1984}
$\mathcal{F}$, defined as $$\mathcal{F}=\frac{\left(\sum_{n=1}^N\omega_n\mathcal{R}_n\right)^2}{N\sum_{n=1}^N(\omega_n\mathcal{R}_n)^2},$$is
employed in our investigation, where $\mathcal{R}_n$ is the rate
of the $n$-th candidate-UE defined in \eqref{eq_r_0} and $\omega_n$
is the probability of the $n$-th candidate-UE being selected to be
served at each coherence slot.

\begin{figure}[t]
\begin{centering}
\includegraphics[width=7.5cm]{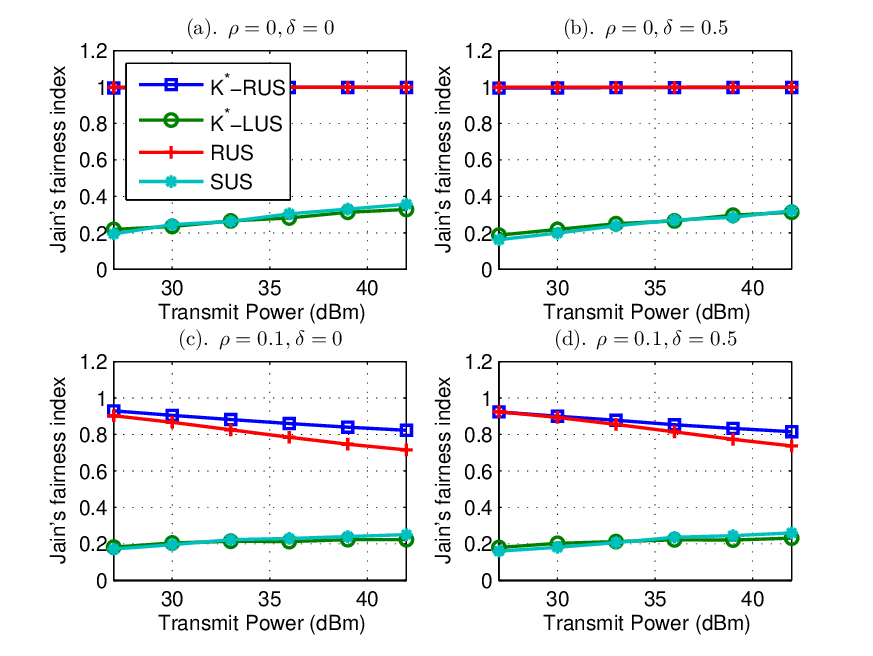}
\par\end{centering}

\centering{}\protect\caption{Fairness performance $\mathcal{F}$ vs. transmit
power $P$. $M=32\text{ and }N=64.$}
\label{fig9}
\end{figure}

\begin{figure}[t]
\begin{centering}
\includegraphics[width=7.5cm]{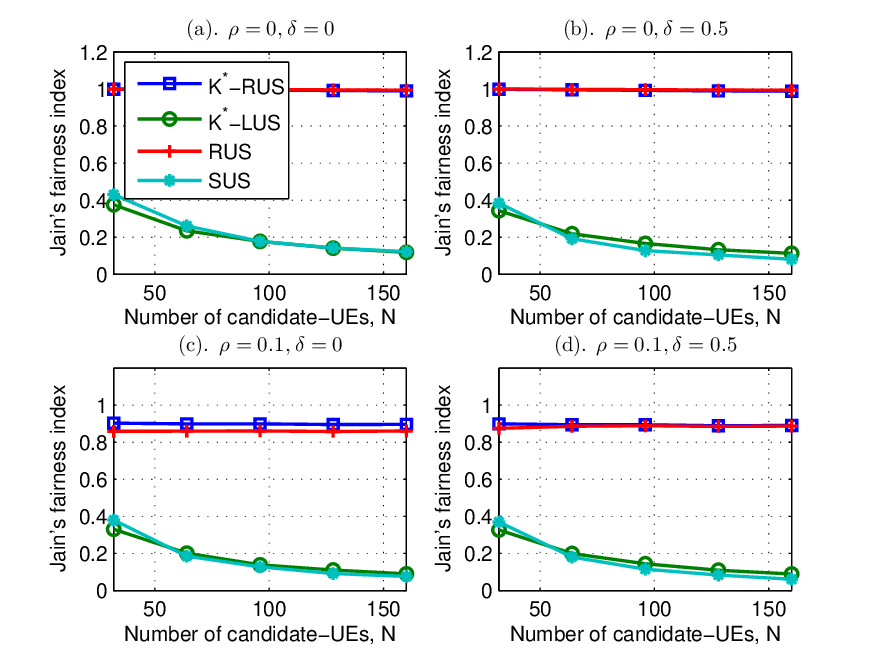}
\par\end{centering}

\centering{}\protect\caption{Fairness performance $\mathcal{F}$ vs. the number
of candidate-UEs $N$. $P=30 \text{ dBm and }M=32.$}
\label{fig10}
\end{figure}

First, we briefly analyze the long-term fairness performance of $K^*$-RUS
and $K^*$-LUS without considering the channel estimation error and
transmit correlation. Then the simulation results are shown in Fig.
\ref{fig9} and Fig. \ref{fig10}.

In the case of $\rho=0,\delta=0$, for the $n$-th candidate-UE in
$K^*$-RUS, we have $\omega_n=K^*_{\text{RUS}}/N$ and $\mathcal{R}_n=\mathcal{R}^*_{\text{RUS}}/K^*_{\text{RUS}}$.
Therefore, the JFI of $K^*$-RUS is given by \begin{equation}\label{eq_jfi1}\mathcal{F}_{\text{RUS}}=\frac{\left(\sum_{n=1}^N\frac{K^*_{\text{RUS}}}{N}\frac{\mathcal{R}_{\text{RUS}}^*}{K^*_{\text{RUS}}}\right)^2}{N\sum_{n=1}^N\left(\frac{K^*_{\text{RUS}}}{N}\frac{\mathcal{R}_{\text{RUS}}^*}{K^*_{\text{RUS}}}\right)^2}=1.\end{equation}It
is clear that $K^*$-RUS is capable of offering the optimal fairness
among candidate-UEs. For $K^*$-LUS, the $K^*_{\text{LUS}}$ candidate-UEs
near to the BS are always active, whereas the other UEs far from the
BS have little chance to be served. Thus, for the candidate-UE which
is the $i$-th nearest to the BS, we have $\omega_i=1,\mathcal{R}_i=\mathcal{R}^*_{\text{LUS}}/K^*_{\text{LUS}},1\leq i\leq K^*_{\text{LUS}}$,
and $\omega_i=0,\mathcal{R}_i=0,K^*_{\text{LUS}}<i\leq N$. The JFI
of $K^*$-LUS is then calculated as \begin{equation}\label{eq_jfi2}\mathcal{F}_{\text{LUS}}=\frac{\left(\sum_{n=1}^{K^*_{\text{LUS}}}\frac{\mathcal{R}_{\text{LUS}}^*}{K^*_{\text{LUS}}}\right)^2}{N\sum_{n=1}^{K^*_{\text{LUS}}}\left(\frac{\mathcal{R}_{\text{LUS}}^*}{K^*_{\text{LUS}}}\right)^2}=\frac{K^*_{\text{LUS}}}{N}.\end{equation}Usually,
we have $\mathcal{F}_{\text{LUS}}<1$ because of $K^*_{\text{LUS}}<N$.
Moreover, we can see that $\mathcal{F}_{\text{LUS}}$ increases when
$K^*_{\text{LUS}}$ becomes higher.

In Fig. \ref{fig9} we show the fairness among UEs against $P$ in
the context of various schemes with $M=32,N=64$. In this simulation,
we assume that the LSF CSI of each candidate-UE remains unchanged
for $100T$ and the SSF CSI changes for each $T$. The window length
evaluated for JFI is also assumed to be $100T$ for creating the worst
scenario in terms of fairness. We can see that $K^*$-RUS provides
good fairness among UEs, and $K^*$-LUS exhibits poor fairness performance.
With perfect channel estimation, the JFI of $K^*$-RUS maintains $1$
for various $P$, which is consistent with \eqref{eq_jfi1}. Furthermore,
for $K^*$-LUS, the JFI increases as $P$ increases. This is because
with the increase of $P$, more UEs can be served at the same time
while $N$ keeps unchanged, which coincides with \eqref{eq_jfi2}.
Note that we also give the simulation results concerning the fairness
performance with imperfect channel estimation. In this context, the
JFI of $K^*$-RUS decreases as $P$ increases. This is because the
SINRs of active-UEs are different from each other due to the existence
of the second part of the denominator of \eqref{eq_sinr_0} (i.e.,
the part related to the channel estimation error). Moreover, the SINR
difference increases when $P$ rises. As a result, the fairness performance
decreases.

Finally, the fairness performance against $N$ for $M=32,P=30$ dBm
is shown in Fig. \ref{fig10}. It can be observed that $K^*$-RUS
is capable of achieving good fairness performance, whereas $K^*$-LUS
has poor fairness performance. Moreover, the increasing of $N$ degrades
the fairness performance of $K^*$-LUS, because the proportion of
active-UEs in candidate-UEs declines although $K^*_{\text{LUS}}$
increases as $N$ becomes larger.

\section{Conclusions}

Considering the requirements of high energy efficiency and massive
device connectivity in the future 5G communication systems, we have
proposed a pair of low-complexity user selection methods for downlink
massive MIMO systems in this paper. Taking the randomness of both
the channel matrix and UE locations into consideration, we have obtained
the approximations of the ergodic sum rates for the multiuser massive
MIMO systems. By exploiting these approximations, $K^*$-RUS and $K^*$-LUS
are developed, which are capable of significantly enhancing the system
sum rate performance. Since no online operations related to SSF CSI
are required in the proposed user selection algorithms, the computational
complexity of the proposed schemes is extremely low. Besides the sum
rate improvements, we also investigated the fairness among UEs and
showed the remarkable fairness performance advantages of the proposed
$K^*$-RUS scheme. In the future, we will investigate low-complexity
user selection methods in multi-cell scenarios.

\section*{Appendix\\Detailed derivation of \eqref{eq_sinr_1}}

With \eqref{eq_ce_0} and \eqref{eq_pc_0}, we have \begin{equation}
\gamma^2=\frac{(1-\rho)P}{\text{tr}(\mathbf{D}^{1/2}\mathbf{Z}_1\mathbf{R}\mathbf{Z}_1^H\mathbf{D}^{1/2})^{-1}}.
\end{equation}According to the results in \cite[Appendix III]{wanger_large_2012},
we obtain $$\text{tr}(\mathbf{D}^{1/2}\mathbf{Z}_1\mathbf{R}\mathbf{Z}_1^H\mathbf{D}^{1/2})^{-1}\xrightarrow{a.s.}\frac{1}{\phi M}\text{tr}(\mathbf{D}^{-1}),$$where
$\phi$ is the unique solution of \begin{equation}\label{eq_rmt_0}
\phi=\frac{1}{M}\text{tr}\left(\mathbf{R}\left(\mathbf{I}_M+\frac{K}{M}\frac{1}{\phi}\mathbf{R}\right)^{-1}\right).
\end{equation}Hence, the deterministic value of $\gamma^2$ satisfies \begin{equation}\label{eq_rmt_1}\gamma^2\xrightarrow{a.s.}\frac{(1-\rho)Pc\phi M}{\sum_{k=1}^Kd_k^{\alpha}}.\end{equation}Then,
we evaluate the second part of the denominator in \eqref{eq_sinr_0}.
Since the entries of $\tilde{\mathbf{h}}_k$ is independent of $\hat{\mathbf{G}}$,
according to \cite[Theorem 3.4]{couillet_random_2011}, we have \begin{equation}\label{eq_rmt_2}\begin{split}
&\tilde{\mathbf{h}}_k\mathbf{R}^{1/2}\hat{\mathbf{G}}^{\dagger}\mathbf{s}\mathbf{s}^H(\hat{\mathbf{G}}^{\dagger})^H(\mathbf{R}^{1/2})^H\tilde{\mathbf{h}}_k^H\\
&\xrightarrow{a.s.}\rho\text{tr}\left(\mathbf{R}\hat{\mathbf{G}}^H(\hat{\mathbf{G}}\hat{\mathbf{G}}^H)^{-2}\hat{\mathbf{G}}\right).
\end{split}\end{equation}Applying \cite[Theorem 14.3]{couillet_random_2011}, we can obtain
$$
\text{tr}\left(\mathbf{R}\hat{\mathbf{G}}^H(\hat{\mathbf{G}}\hat{\mathbf{G}}^H)^{-2}\hat{\mathbf{G}}\right)\xrightarrow{a.s.}\frac{1}{1-\rho}\frac{\psi}{\frac{M}{K}\phi^2-\psi}\frac{1}{K}\text{tr}(\mathbf{D}^{-1}),
$$where $\phi$ is given by \eqref{eq_rmt_0} and $\psi$ is defined
as $$\psi=\frac{1}{M}\text{tr}\left(\mathbf{R}^2\left(\mathbf{I}_M+\frac{K}{M}\frac{1}{\phi}\mathbf{R}\right)^{-2}\right).$$Thus,
for \eqref{eq_rmt_2}, we obtain \begin{equation}\label{eq_rmt_3}\begin{split}
&\tilde{\mathbf{h}}_k\mathbf{R}^{1/2}\hat{\mathbf{G}}^{\dagger}\mathbf{s}\mathbf{s}^H(\hat{\mathbf{G}}^{\dagger})^H(\mathbf{R}^{1/2})^H\tilde{\mathbf{h}}_k^H\\
&\xrightarrow{a.s.}\frac{\rho}{1-\rho}\frac{\psi}{\frac{M}{K}\phi^2-\psi}\frac{\sum_{k=1}^Kd_k^{\alpha}}{cK}.
\end{split}\end{equation}Therefore, substituting \eqref{eq_rmt_1} and \eqref{eq_rmt_3} into
\eqref{eq_sinr_0}, we obtain \eqref{eq_sinr_1}.

\begin{IEEEbiography}[{\includegraphics[width=1in,height=1.25in,clip,keepaspectratio]{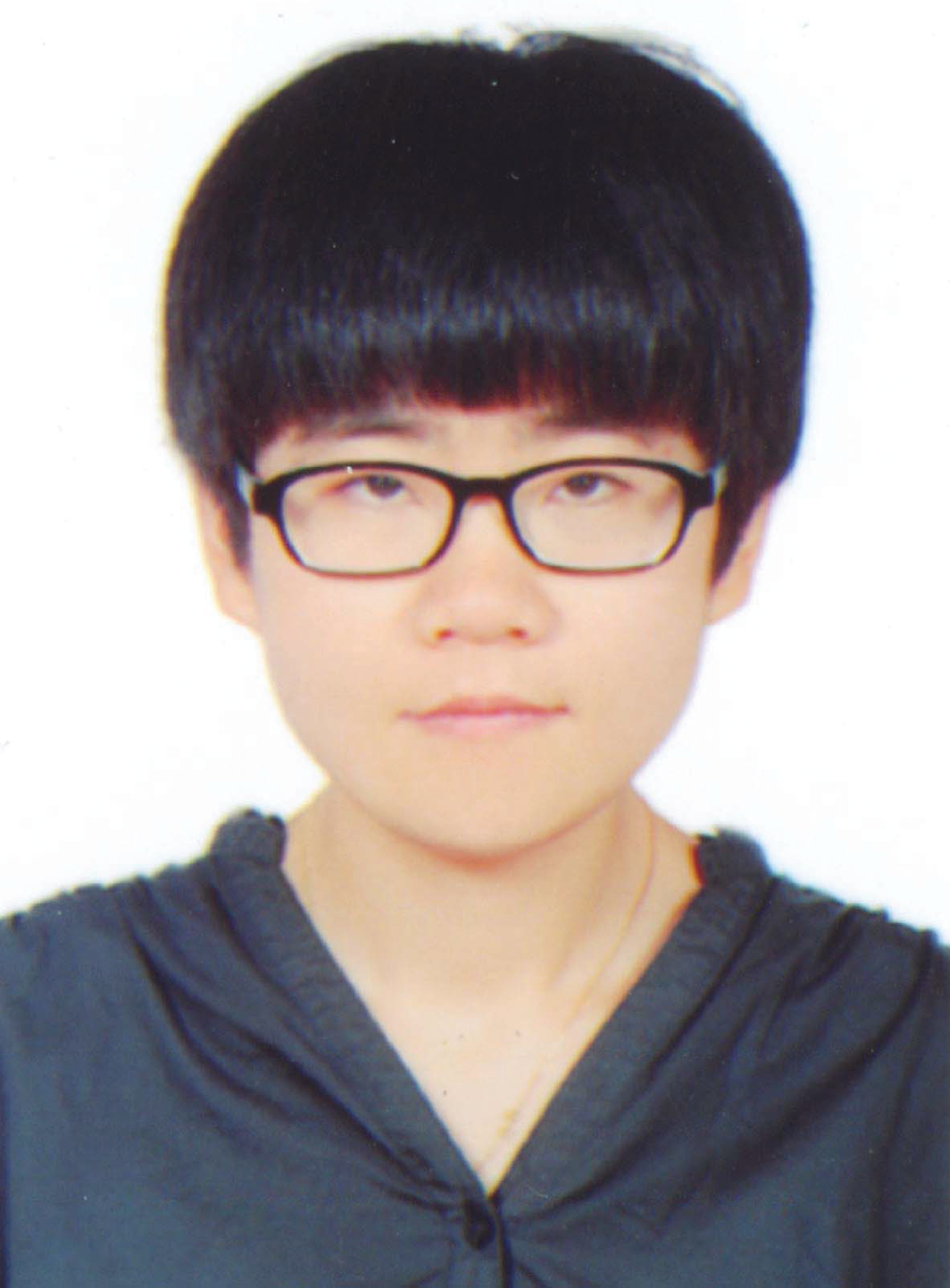}}]{Haijing Liu}
received her B. S. and M.S. degrees in  Communication and Information System from Southeast University, Nanjing, China, in 2007 and 2010, respectively.
From 2010 to 2012, she worked in Alcatel-Lucent Shanghai Bell Co. Ltd as an hardware engineer.
She is currently pursuing the Ph.D. degree at Beijing University of Posts and Telecommunications, Beijing, China.
Her research interests include massive MIMO systems and low-complexity digital signal processing in wireless communications.
\end{IEEEbiography}
\begin{IEEEbiography}[{\includegraphics[width=1in,height=1.25in,clip,keepaspectratio]{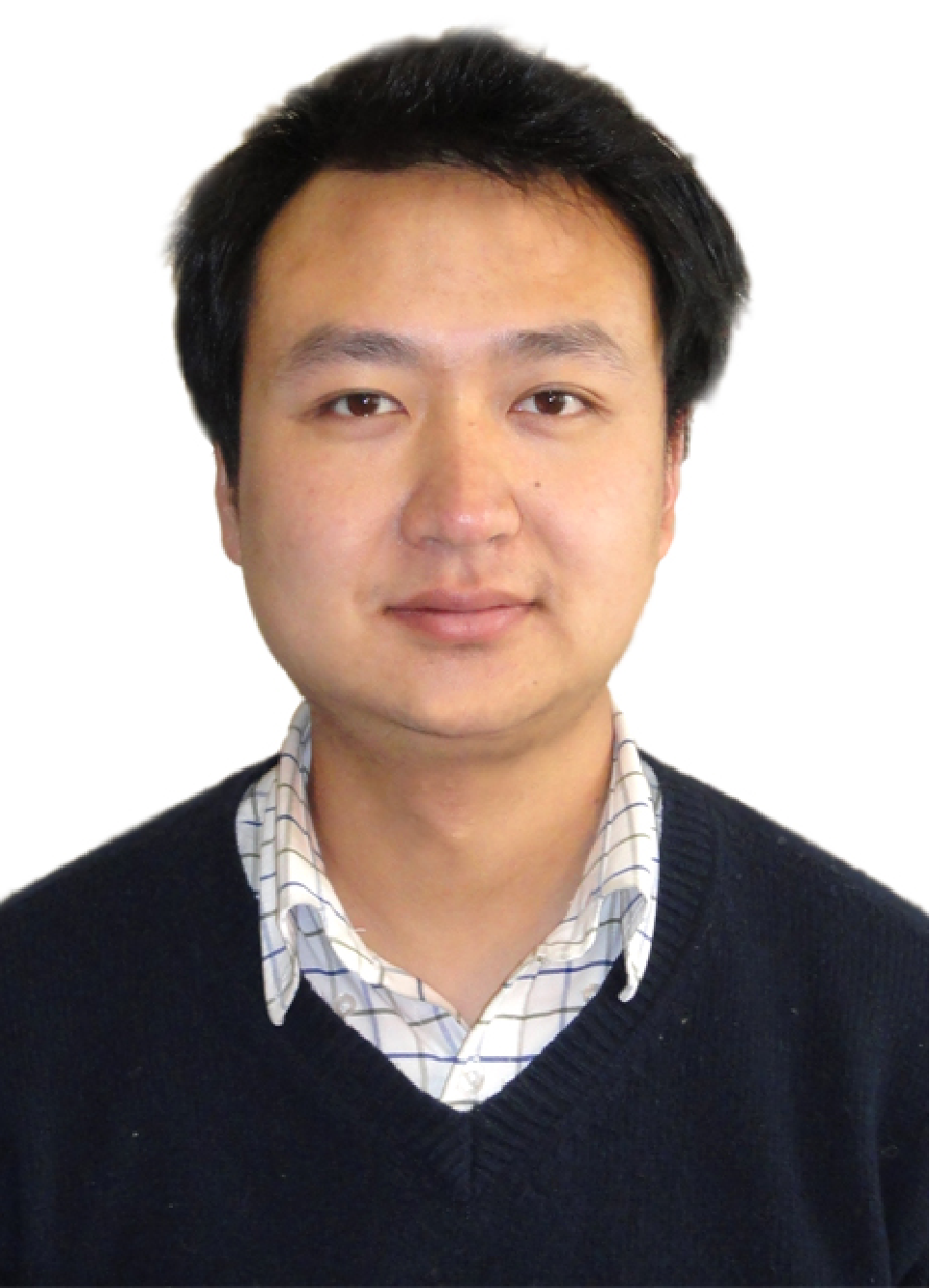}}]{Hui Gao}
(S'10-M'13) received the B. Eng. degree in information engineering and the Ph.D. degree in signal and information processing from Beijing University of Posts and Telecommunications (BUPT), Beijing, China, in July 2007 and July 2012, respectively. From May 2009 to June 2012, he also served as a Research Assistant for the Wireless and Mobile Communications Technology R$\&$D Center, Tsinghua University, Beijing, China. From April 2012 to June 2012, he visited Singapore University of Technology and Design (SUTD), Singapore, as a Research Assistant. From July 2012 to February 2014, he was a Postdoc Researcher with SUTD. He is now with the School of Information and Communication Engineering, BUPT, as an Assistant Professor. His research interests include massive MIMO systems, cooperative communications, ultra-wideband wireless communications.
\end{IEEEbiography}
\begin{IEEEbiography}[{\includegraphics[width=1in,height=1.25in,clip,keepaspectratio]{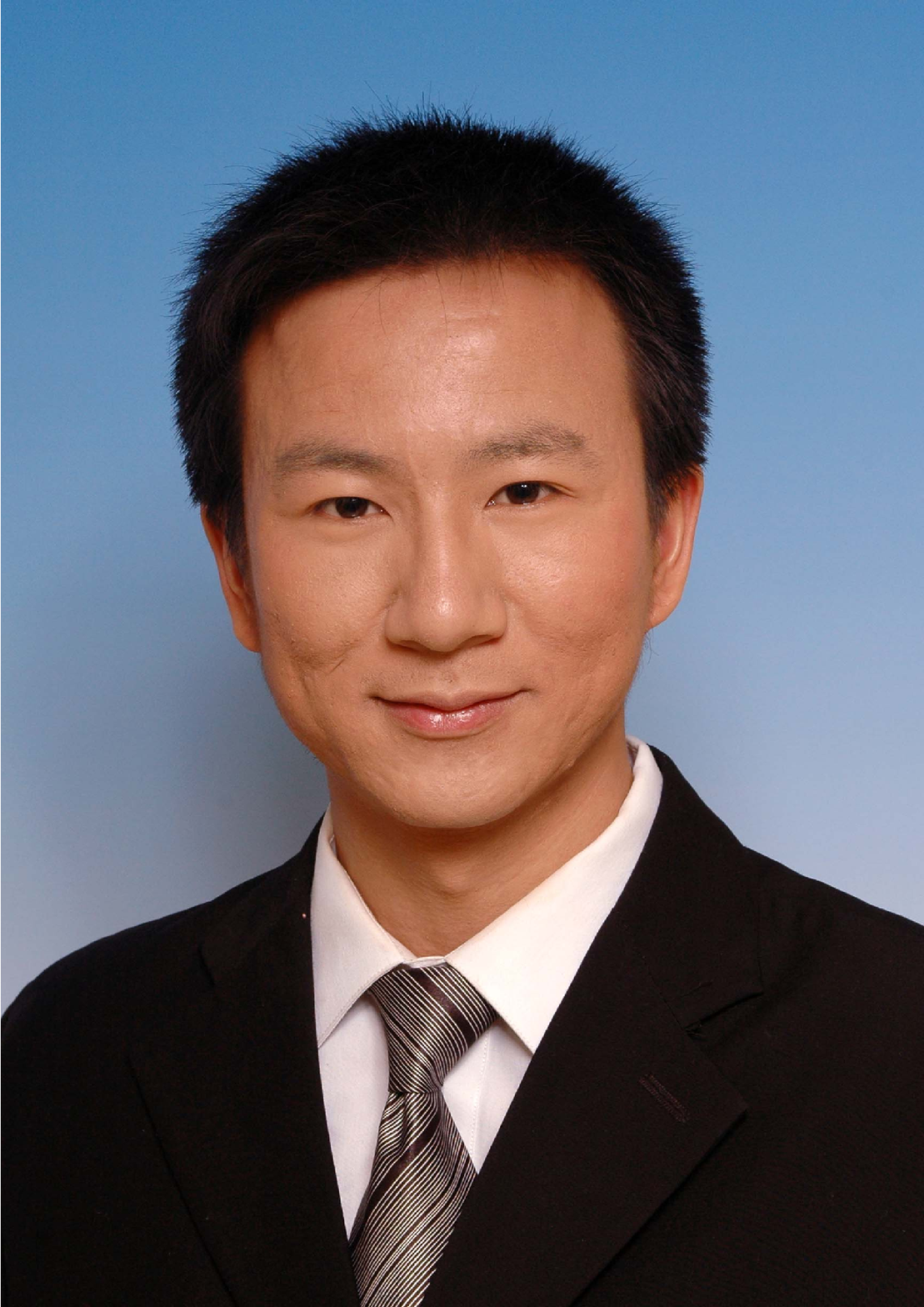}}]{Shaoshi Yang}
(S'09-M'13)  received
the B.Eng. Degree in Information Engineering
from Beijing University of Posts and
Telecommunications (BUPT), China, in 2006, the first Ph.D. Degree in Electronics and Electrical Engineering from University of Southampton, U.K., in 2013, and a second Ph.D. Degree in Signal and Information Processing from BUPT in 2014. Since 2013 he has been a Postdoctoral Research Fellow in
University of Southampton, U.K., and from 2008 to 2009, he was
an Intern Research Fellow with the Intel Labs China, Beijing, where he focused on
Channel Quality Indicator Channel design for mobile WiMAX (802.16 m).
His research interests include MIMO signal processing, green radio, heterogeneous networks, cross-layer interference management, convex optimization and its applications. He has published in excess of 30 research papers on IEEE journals and conferences.

Shaoshi has received a number of academic and research awards, including the PMC-Sierra Telecommunications Technology Scholarship at BUPT, the Electronics and Computer Science (ECS) Scholarship of University of Southampton and the Best PhD Thesis Award of BUPT. He serves as a TPC member of a number of IEEE conferences and journals, including \textit{IEEE ICC, PIMRC, ICCVE, HPCC} and \textit{IEEE Journal on Selected Areas in Communications}. He is also a Junior Member of the Isaac Newton Institute for Mathematical Sciences, Cambridge University, UK. (https://
sites.google.com/site/shaoshiyang/)
\end{IEEEbiography}
\begin{IEEEbiography}[{\includegraphics[width=1in,height=1.25in,clip,keepaspectratio]{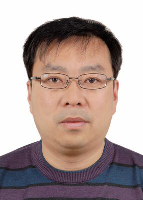}}]{Tiejun Lv}
(M'08-SM'12) received the M.S. and
Ph.D. degrees in electronic engineering from the
University of Electronic Science and Technology
of China (UESTC), Chengdu, China, in 1997 and
2000, respectively. From January 2001 to December
2002, he was a Postdoctoral Fellow with Tsinghua
University, Beijing, China. From September 2008
to March 2009, he was a Visiting Professor with
the Department of Electrical Engineering, Stanford
University, Stanford, CA, USA. He is currently a Full
Professor with the School of Information and Communication
Engineering, Beijing University of Posts and Telecommunications
(BUPT). He is the author of more than 200 published technical papers on
the physical layer of wireless mobile communications. His current research
interests include signal processing, communications theory and networking.
He was the recipient of the Program for New Century Excellent Talents in
University Award from the Ministry of Education, China, in 2006.
\end{IEEEbiography}

\end{document}